\documentclass[twocolumn,aps,prd,showpacs,preprintnumbers,amsmath,amssymb,showpacs,showkeys,nofootinbib]{revtex4}
\usepackage{latexsym,epsfig,amsmath,amssymb}
\usepackage{amssymb}
\usepackage{amsmath}
\usepackage{amsfonts}
\usepackage{epsfig} 
\usepackage{verbatim} 

\newcommand{\seq}{\begin{subequations}}
\newcommand{\sen}{\end{subequations}}
\newcommand{\eq}{\begin{eqnarray}}
\newcommand{\en}{\end{eqnarray}}

\def\Lfv{$L_f\hspace{-0.95em}/\ \ $}

\begin{document}

\title{Lepton flavor violating decays of vector mesons} 

\noindent
\author{Thomas Gutsche$^1$, 
        Juan C. Helo$^2$,        
        Sergey Kovalenko$^2$,        
        Valery E. Lyubovitskij$^1$\footnote{On leave of absence
        from Department of Physics, Tomsk State University,
        634050 Tomsk, Russia}
\vspace*{1.2\baselineskip}}

\affiliation{$^1$ Institut f\"ur Theoretische Physik,  
Universit\"at T\"ubingen,\\
Kepler Center for Astro and Particle Physics, \\ 
Auf der Morgenstelle 14, D--72076 T\"ubingen, Germany
\vspace*{1.2\baselineskip} \\
$^2$ Centro de Estudios Subat\'omicos, 
Universidad T\'ecnica Federico Santa Mar\'\i a, \\
Centro Cient\'{i}fico-T\'ecnologico de  Valpara\'{i}so,\\
Casilla 110-V, Valpara\'\i so, Chile\\}

\date{\today}

\begin{abstract} 

We estimate the rates of lepton flavor violating decays 
of the vector mesons $\rho, \omega, \phi \to e \mu$.
The theoretical tools are based
on an effective Lagrangian approach without referring to any specific 
realization of the physics beyond the standard model responsible 
for lepton flavor violation (\Lfv). The effective lepton-vector meson 
couplings are extracted from the existing experimental bounds 
on the nuclear $\mu^--e^-$ conversion. 
In particular, we  derive an upper limit for the \Lfv branching ratio 
of $\phi$ mesons with  
${\rm Br}(\phi \to e \mu) = 1.3 \times 10^{-21}$ which is much more stringent
than the recent experimental result 
${\rm Br}(\phi \to e \mu) < 2 \times 10^{-6}$ 
presented by the SND Collaboration.  Our derived, very tiny limits on 
\Lfv decays of vector mesons clearly prevent a possible, direct experimental 
observation of these processes.
 
\end{abstract}



\maketitle

The study of lepton flavor violation (\Lfv) processes involving
charged leptons is an important tool to search for New Physics 
beyond the Standard Model (SM). Recently the SND Collaboration at the 
BINP (Novosibirk)~\cite{Achasov:2009en} reported on the 
search for the \Lfv process $e^+e^-\to e\mu$ in the energy region 
$\sqrt{s}=984 - 1060$ MeV at the VEPP-2M $e^+e^-$ 
collider. They give a model independent upper limit on the 
$\phi\to e\mu$ branching fraction of
${\rm Br}(\phi\to e\mu) < 2 \times 10^{-6}$.
In the literature there already exists a stringent limit with
${\rm Br}(\phi\to e\mu) \le 4 \times 10^{-17}$~\cite{Nussinov:2000nm}, 
deduced from  the existing experimental bounds on the \Lfv $\mu \to 3e$ decay. 

In Refs.~\cite{Faessler:2004jt,Faessler:2005hx} we 
studied the \Lfv process of nuclear $\mu^--e^-$ conversion in nuclei.
This work was set up within the general framework 
of an effective Lagrangian approach without referring to any specific 
realization of physics beyond the SM responsible for \Lfv. 
We examined the impact of specific hadronization prescriptions on 
New Physics contribution to nuclear $\mu^--e^-$ conversion and stressed
the importance of vector and scalar meson exchange between lepton and 
nucleon currents. As one consequence we 
derived limits on various \Lfv couplings of vector mesons to
$\mu-e$ current using existing 
experimental data on $\mu^{-}-e^{-}$ conversion in nuclei. The purpose of 
the present letter is to apply these limits to set upper bounds on the 
rates of the vector meson decays $\rho, \omega, \phi \to e \mu$. 

The contribution of vector mesons to $\mu^{-}-e^{-}$ conversion 
in nuclei is shown 
in~Fig.~1. 
\begin{figure}
\begin{center} 
\epsfig{file=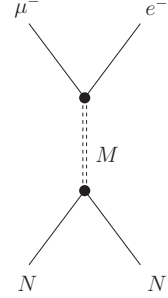, scale=.7} 
\end{center} 
\vspace*{-.75cm}
\caption{Contribution of vector mesons to nuclear $\mu^{-}-e^{-}$ conversion.} 
\end{figure} 
In this diagram the upper vertex corresponds to the \Lfv 
interactions of  vector mesons $M = \rho, \omega, \phi$ with $e, \mu$ 
given by  the following model--independent Lagrangian~\cite{Faessler:2004jt}: 
\eq\label{eff-LV}
{\cal L}_{eff}^{lM} =   ( \xi_V^{M} j_{\mu}^V\ + \xi_A^{M} 
j_{\mu}^A\ )M^{\mu} \,,
\en 
where $M=\rho,\omega,\phi $ and $\xi_{V,A}^{M} $ 
are effective vector and axial 
couplings of a vector meson $M$ to the \Lfv  lepton currents 
$j_{\mu}^V = \bar e \gamma_{\mu} \mu$ and  
$j_{\mu}^A = \bar e \gamma_{\mu} \gamma_{5} \mu$.
The lower vertex of the diagram in Fig 1 is described by the 
conventional nucleon-vector meson Lagrangian of the 
form~\cite{Weinberg:de,Mergell:1995bf,Kubis:2000zd}: 
\begin{eqnarray}\label{MN} 
{\cal L}_{MNN} = \frac{1}{2}  
\bar{N}\gamma^{\mu}\left[ g_{_{\rho NN}} \, \vec{\rho}_{\mu} \, 
\vec{\tau} + g_{_{\omega NN}} \, \omega_{\mu} +  
g_{_{\phi NN}} \, 
\phi_{\mu}\right] N . 
\end{eqnarray} 
In this Lagrangian we neglected the derivative terms which are irrelevant for 
coherent $\mu^- - e^-$ conversion. For the meson-nucleon couplings 
$g_{MNN}$ we use the numerical values taken from an updated dispersive 
analysis~\cite{Mergell:1995bf,Meissner:1997qt} 
\begin{eqnarray}\label{VN-couplings}
g_{_{\rho NN}}= 4.0\,, \,\, 
g_{_{\omega NN}} = 41.8\,, \,\, 
g_{_{\phi NN}} = - 0.24\,. 
\end{eqnarray} 
Starting from the Lagrangian  ${\cal L}_{eff}^{M} = {\cal L}_{eff}^{lM} 
+ {\cal L}_{MNN}$ 
of Eqs.~(\ref{eff-LV}) and (\ref{MN}) it is straightforward to 
derive the contribution of the diagram in Fig. 1 to the total $\mu^- - e^-$ 
conversion branching ratio \cite{Faessler:2004jt}. To the leading order 
of the non-relativistic reduction the coherent $\mu^--e^-$ conversion 
branching ratio takes the form~\cite{Kosmas:ch} 
\begin{equation} 
R_{\mu e^-}^{coh} \ = \  
\frac{{\cal Q}} {2 \pi } \  \   
\frac{p_e E_e } 
{ \Gamma ({\mu^-\to {\rm capture}}) } 
\, , 
\label{Rme}
\end{equation} 
where $p_e, E_e$ are 3-momentum and energy of the outgoing electron 
(for details see Ref.~\cite{Faessler:2004jt,Faessler:2005hx});
$\Gamma ({\mu^-\to {\rm capture}})$ is the total rate of the ordinary
muon capture reaction.
The factor ${\cal Q}$ in Eq.~(\ref{Rme}) has the 
form~\cite{Faessler:2004jt,Kosmas:2001mv} 
\begin{eqnarray}
\nonumber
\hspace*{-1cm}
{\cal Q} &=& |({\cal M}_{p} + {\cal M}_{n})\alpha_{VV}^{(0)}+ 
({\cal M}_{p} - {\cal M}_{n})\alpha_{VV}^{(3)} |^2 +\\
&+&|({\cal M}_{p} + {\cal M}_{n})\alpha_{AV}^{(0)}+
({\cal M}_{p} - {\cal M}_{n})\alpha_{AV}^{(3)} |^2 \, . 
\label{Rme.1} 
\end{eqnarray} 
It contains the nuclear matrix elements ${\cal M}_{p,n}$ which have been
calculated numerically in Refs.~\cite{Kosmas:2001mv} for various nuclei. 
Here we consider $\mu^--e^-$ conversion in ${}^{48}$Ti studied by
the SINDRUM II Collaboration  \cite{Honecker:zf}. For this nucleus
we have ${\cal M}_{p}\approx 0.104, {\cal M}_{n}\approx 0.127$.
The  ${\cal Q}$ factor also contains the \Lfv lepton-nucleon parameters
$\alpha_{VV,AV}$.
For the contribution of the meson-exchange diagram in Fig. 1 
these coefficients are expressed in terms of the \Lfv couplings 
$\xi_{V,A}^{\rho,\omega,\phi}$ of Eq. (\ref{eff-LV}) 
as~\cite{Faessler:2004jt,Faessler:2005hx}
\begin{eqnarray} \label{alpha-V-ex}
\alpha_{aV}^{(3)} &=& - \frac{1}{2} 
\frac{g_{_{\rho NN}}}{m_{\rho}^2 
+ m_{\mu}^2} \xi_{a}^{\rho},
\\   
\nonumber
\alpha_{aV}^{(0)}&=& -\frac{1}{2} 
\frac{g_{_{\omega NN}}}{m_{\omega}^2 
+ m_{\mu}^2} \xi_{a}^{\omega} 
- \frac{1}{2} 
\frac{g_{_{\phi NN}}}{m_{\phi}^2 
+ m_{\mu}^2} \xi_{a}^{\phi}\,,
\end{eqnarray}
with $a=V,A$. Here $m_{\rho}, m_{\omega}, m_{\phi}$ and $m_{\mu}$ are 
the vector meson and muon masses respectively. 
In Ref.~\cite{Faessler:2004jt} we extracted upper limits on the couplings 
$\alpha_{aV}^{(i)}$ from the experimental upper bounds on $\mu^{-}-e^{-}$ 
conversion in ${}^{48}$Ti reported by the SINDRUM II 
Collaboration~\cite{Honecker:zf}. These limits can be translated into 
limits on the \Lfv couplings $\xi_{V,A}^{\rho,\omega,\phi}$. Assuming no 
accidental cancellations between different terms in (\ref{alpha-V-ex}) 
and using the values of the meson-nucleon couplings  
from Eq. (\ref{VN-couplings}) we get 
\begin{eqnarray}\label{limits-on-LFV-couplings}
\hspace*{-0.2cm}
\xi_{a}^{\rho} \leq 3.6 \times 10^{-12}, \  \xi_{a}^{\omega} 
\leq 3.6 \times 10^{-14}, \  \xi_{a}^{\phi} \leq 1.0 \times 10^{-11}.
\end{eqnarray}

We note that the Lagrangian (\ref{eff-LV}) also governs the \Lfv decay 
of vector mesons $M \to e + \mu$. Thus, using the limits 
from Eq.  (\ref{limits-on-LFV-couplings}) we can set upper bounds on the rates
of these two-body decays.  
Their branching ratios are given by: 
\vspace*{-5mm}
\eq
{\rm Br}(M \to e + \mu) \simeq 
\frac{(\xi_{V}^{M})^{2}+(\xi_{A}^{M})^{2}}{12\pi \Gamma^{M}_{tot}} \, 
m_M  \, \biggl( 1 - \frac{3}{2} r_M^2 \biggr) \, , 
\en \\
where $r_M = m_\mu/m_M$ with $M=\rho, \omega, \phi$ and $\Gamma^{M}$ 
is the total decay width of meson $M$. Here we neglect the electron mass.  
With the limits set by Eq. (\ref{limits-on-LFV-couplings}) we get the 
following  upper limits on the branching ratios 
of the vector meson \Lfv decays
\vspace*{-3mm}
\begin{eqnarray}\label{Limits-on-branchings}
&&{\rm Br}(\rho \to e + \mu) \leq 3.5 \times 10^{-24}\,,\\
&&{\rm Br}(\omega \to e + \mu) \leq 6.2 \times 10^{-27}\,,\\ 
&&{\rm Br}(\phi \to e + \mu) \leq 1.3 \times 10^{-21}\,. 
\end{eqnarray}
In comparison with the existing limit on the transition $\phi \to e + \mu$, 
extracted from $\mu\rightarrow 3e$ in Ref.~\cite{Nussinov:2000nm}, 
our limit is about 4 orders of magnitude more stringent.
The experimental upper bound on this process
recently reported by the SND Collaboration~\cite{Achasov:2009en} is even 
15 orders of magnitude larger than the one which we 
derived from $\mu^{-}-e^{-}$ conversion.  

{\it In conclusion,}  we extracted from the experimental data on 
nuclear $\mu^{-}-e^{-}$ conversion  new upper limits on the \Lfv couplings 
of vector mesons to the $e-\mu$ lepton currents.  
Then we applied these limits to deduce bounds on the branching ratios of 
$\rho, \omega, \phi \to e + \mu$ decays.
The obtained limits (\ref{Limits-on-branchings}) 
are significantly more stringent 
than those exiting in the literature. In view of these bounds, which 
are 15 orders of magnitude below the current experimental upper limit, 
the prospects of direct experimental observation of 
$\rho, \omega, \phi \to e + \mu$ decays look rather pessimistic. 

\begin{acknowledgments}

This work was supported by the DFG under Contract No. FA67/31-2 
and No. GRK683, the European Community-Research Infrastructure 
Integrating Activity ``Study of Strongly Interacting Matter'' (HadronPhysics2,
Grant Agreement No. 227431), the President grant of Russia
``Scientific Schools''  No. 3400.2010.2, by CONICYT (Chile) via
the projects PBCT ACT-028 and PFB 8208 as well as by Russian Science 
and Innovations Federal Agency under contract  No 02.740.11.0238. 

\end{acknowledgments}


\begin{thebibliography}{99} 
\bibitem{Achasov:2009en}
  M.~N.~Achasov {\it et al.},
  arXiv:0911.1232 [hep-ex]. 
\bibitem{Nussinov:2000nm}
  S.~Nussinov, R.~D.~Peccei and X.~M.~Zhang,
  Phys.\ Rev.\  D {\bf 63}, 016003 (2001)
\bibitem{Faessler:2004jt}  A.~Faessler, T.~Gutsche, 
S.~Kovalenko, V.~E.~Lyubovit\-skij,  
  I.~Schmidt and F.~Simkovic,
  Phys.\ Lett.\  B {\bf 590}, 57 (2004); 
  A.~Faessler, T.~Gutsche, S.~Kovalenko, 
  V.~E.~Lyu\-bo\-vit\-skij, 
  I.~Schmidt and F.~Simkovic,
  Phys.\ Rev.\  D {\bf 70}, 055008 (2004)
\bibitem{Faessler:2005hx}
  A.~Faessler, T.~Gutsche, S.~Kovalenko, V.~E.~Lyubovitskij and I.~Schmidt,
  Phys.\ Rev.\  D {\bf 72}, 075006 (2005)
  \bibitem{Weinberg:de}
S.~Weinberg,
Phys.\ Rev.\  {\bf 166}, 1568 (1968); 
J.~J.~Sakurai, {\it Currents and Mesons}, 
Chicago Lectures in Physics (The University of Chicago Press,  
Chicago and London, New York, 1967) 
%
\bibitem{Mergell:1995bf}
P.~Mergell, U.~G.~Meissner and D.~Drechsel,
Nucl.\ Phys.\ A {\bf 596}, 367 (1996)
%
\bibitem{Kubis:2000zd}
B.~Kubis and U.~G.~Meissner,
Nucl.\ Phys.\ A {\bf 679}, 698 (2001)
%
\bibitem{Meissner:1997qt}
U.~G.~Meissner, V.~Mull, J.~Speth and J.~W.~van Orden,
Phys.\ Lett.\ B {\bf 408}, 381 (1997)
%
\bibitem{Kosmas:ch}
T.~S.~Kosmas, G.~K.~Leontaris and J.~D.~Vergados,
Prog. \ Part. \ Nucl. \ Phys. \ {\bf 33}, 397 (1994) 
%
\bibitem{Kosmas:2001mv}
T.~S.~Kosmas, S.~Kovalenko and I.~Schmidt,
Phys.\ Lett.\ B {\bf 511}, 203 (2001);  
Phys.\ Lett.\ B {\bf 519}, 78 (2001) 
\bibitem{Honecker:zf} W.~Honecker {\it et al.} [SINDRUM II Collaboration],
Phys.\ Rev.\ Lett.\  {\bf 76}, 200 (1996) 

\end{thebibliography}
\end{document}